\documentclass{article}

\usepackage[english]{babel}

\usepackage[letterpaper,top=2cm,bottom=2cm,left=3cm,right=3cm,marginparwidth=1.75cm]{geometry}

\usepackage{amsmath}
\usepackage{graphicx}
\usepackage[colorlinks=true, allcolors=blue]{hyperref}

\begin{document}

\title{An Application of the Ornstein-Uhlenbeck Process to Pairs Trading}

\author{Jirat Suchato, 
       Sean Wiryadi,
       Danran Chen,
       Ava Zhao,
       Michael Yue
       }
\date{}
\maketitle

\section*{Disclaimer}

The results of this paper are preliminary and intended for educational purposes only. The paper was authored as part of coursework for the class Data-Driven Methods in Finance, taught at Columbia University. The content of this paper is not financial advice. The authors disclaim any liability for financial losses or other consequences based on the usage of this paper.

\section*{Abstract}

In this paper, we conduct preliminary analysis on a pairs trading strategy using the Ornstein-Uhlenbeck process to model stock price differences and compare that to a naive pairs trading strategy using a rolling window to calculate mean and standard deviation parameters. Our preliminary findings suggest that running a pairs trading strategy with the Ornstein-Uhlenbeck process outperforms the naive pairs trading strategy on a risk-return basis. Key further research can be conducted on the selection of pairs, augmenting the investment universe, finding different criteria in pairs selection, applying more rigorous machine learning techniques to assist with forecasting pricing trends, and in integrating portfolio optimization techniques.

\section*{Introduction}

\noindent Pairs trading is a widely used market-neutral strategy in quantitative finance that capitalizes on the price convergence of two historically correlated securities. By simultaneously shorting the overpriced asset and longing the underpriced one, the strategy aims to profit from the mean-reverting behavior of their price spread. Unlike directional trading strategies, pairs trading does not depend on the overall market direction, allowing it to be robust in volatile or stagnant market conditions.\newline

\noindent This project seeks to improve the traditional pairs trading framework by leveraging statistical and econometric techniques. Specifically, we explore the application of cointegration tests, and modelling the spread as a mean-reverting processes using the Ornstein-Uhlenbeck (OU) model. The methodology is tested on historical market data to evaluate its performance and robustness under varying market conditions.\newline

\noindent The selection of pairs is critical. Pairs chosen should exhibit high correlation between the percentage change of the prices of each security, and the price spread should exhibit a mean reverting behavior. Once pairs are chosen, an algorithm can be developed to model the spread and to forecast future pricing trends to take a directional view on the spread. In our case, we employ the Ornstein-Uhlenbeck process to do so, and we compare our results to a basic computation of the z score using the rolling mean and standard deviation by backtesting our results on historical pricing data.\newline

\section*{Data}

We defined the universe of stocks to analyze as all US-based companies with a market capitalization greater than \$1 billion. The stock data was obtained from Refinitiv's Datastream, specifically from the \textit{wrds\_ds2dsf} table, which provides financial metrics such as market capitalization, returns, and prices. We limited the dataset to companies listed in the US, trading in USD, and filtered for data starting from 2018. This allowed us to focus on liquid, large-cap stocks for the analysis.\newline

\noindent Additionally, we retrieved metadata from the \textit{wrds\_ds\_names} table, which includes key identifiers like company names, tickers, and other relevant securities information. After obtaining this data, we removed duplicates based on the combination of \textit{infocode} and \textit{dscode}, ensuring that each security was represented uniquely.\newline

\noindent The next step was merging the filtered stock data (the universe) with the metadata. By performing an inner join on both \textit{infocode} and \textit{dscode}, we ensured that only the securities present in both datasets were retained. This merge allowed us to combine financial data with corresponding company names and other identifiers, creating a comprehensive dataset. Finally, an additional merge was performed to include the general industry descriptions and specific industry group descriptions. The final dataset contains 3,986,371 rows and 26 columns.\newline

\noindent After merging, we selected a subset of important columns that were relevant to our analysis. These columns included essential identifiers such as \textit{infocode}, \textit{dscode}, \textit{ticker}, and \textit{isin}, as well as financial metrics like market capitalization, returns, and closing prices and industry descriptions. An example of the data is shown in Table~\ref{tab:table1} and Table~\ref{tab:table2}.

\section*{Pair Selection}

\noindent To quantify the similarity between the returns of two companies, the squared mean squared distance (MSD) was computed for each pair. This was calculated only on data that was from 2018 to 2020 to ensure point in time analysis. The data which the pairs trading strategy was analyzed on was after the 2018-2020 period. The mean squared distance between two companies \( i \) and \( j \), based on their returns $d_{ij}$, is defined as:
\begin{equation}
    d_{ij} = \frac{1}{N} \sum_{t=1}^{N} \left( R_{i,t} - R_{j,t} \right)^2
    \label{eq:MSE}
\end{equation}
where \( R_{i,t} \) and \( R_{j,t} \) represent the returns of company \( i \) and company \( j \) at time \( t \), respectively.\newline

\noindent The pairs with the smallest pairwise MSD values were selected, as we assume they demonstrated the highest degree of return similarity. To ensure independence and avoid redundancy, once a company was included in a pair, it was excluded from further pairings. This process resulted in a list of unique pairs with minimal return deviations, serving as strong candidates for further validation.

\section*{Pair Validation}

Following the distance-based selection, the statistical relationship between the returns of each pair was tested using the Engle-Granger cointegration test. The cointegration test was performed by running an Ordinary Least Squares (OLS) regression on the returns of one company against the other and analyzing the residuals for stationarity. The test produced a cointegration statistic and a corresponding p-value, where a p-value below 0.05 indicated a statistically significant long-term equilibrium relationship between the two time series. Pairs that passed this cointegration test were retained as valid trading candidates, as their returns exhibited mean-reverting behavior—a crucial property for pairs trading strategies. Finally, a manual review of the pairs was conducted, evaluating company characteristics such as industry and supplier relationships. Based on this review, the pairs were categorized into three groups: great, average, and poor. This three-step process—combining distance-based ranking, cointegration testing, and manual evaluation—ensured that the final selected pairs not only had similar return profiles but also demonstrated stable long-term relationships suitable for trading.\newline

\noindent For instance, the pair \textbf{United Airlines} and \textbf{Delta Airlines} returned a cointegration test statistic of \(-12.0389\) with a p-value of \(0.000\), demonstrating a statistically significant relationship suitable for trading. In contrast, the pair \textbf{Credit Corp LTD} and \textbf{Farmers \& Merchants Bank} produced a cointegration test statistic of \(-0.252617\) with a p-value of \(0.9779\), indicating no significant cointegration and, therefore, making it unsuitable for further analysis.

\section*{Trading Strategy}

\subsection*{Baseline Strategy}

The baseline model is a naive pairs trading algorithm, which computes the rolling mean and standard deviation for the spread over a 30-day period, $\mu$ and $\sigma$ where:

\begin{equation}
    \mu_{ij} = \frac{1}{30} \sum_{t = -30}^{0}S_{ijt}
\end{equation}

\begin{equation}
    \sigma_{ij} = \sqrt{\frac{(S_{ij}-\mu_{ij})^2}{30-1}}
\end{equation}

\noindent This model serves as a baseline, it is common practice for a standard pairs trading strategy computes a z-score using the rolling mean and standard deviation.

\subsection*{OU Model Integration}
The pairs trading strategy begins by calculating the price spread between two assets.
For each time period t, the spread between stock i and stock j was calculated as

\begin{equation}
    S_{ijt} = P_{it} - P_{jt}
\end{equation}

\noindent The baseline model is a naive pairs trading algorithm, which computes the rolling mean and standard deviation for the spread over a 30-day period, $\mu$ and $\sigma$ where:

\begin{equation}
    \mu_{ij} = \frac{1}{30} \sum_{t = -30}^{0}S_{ijt}
\end{equation}

\noindent This spread is assumed to follow an Ornstein-Uhlenbeck (OU) process, which captures mean-reverting behavior.\newline

\noindent The Ornstein-Uhlenbeck process is governed by the stochastic differential equation

\begin{equation}
    dS_{t} = \lambda (\mu - S_{t})dt + \sigma dW_{t}
\end{equation}

\noindent which can be discretized as

\begin{equation}
    S_{t + 1} = S_{t} e^{-\lambda \delta} + \mu (1 - e^{-\lambda \delta}) + 
    \sigma \sqrt{\frac{1 - e^{2 \sigma \delta}}{2 \lambda}} N_{0,1}
\end{equation}

\noindent The discrete form of the SDE can be written into linear regression form

\begin{equation}
    S_{t + 1} = a S_{t} + b + \epsilon
    \label{eq:linreg}
\end{equation}

\noindent where

\begin{equation}
    a = e^{-\lambda \delta}; \quad b = \mu (1 - e ^{-\lambda \delta}); \quad \text {sd} (\epsilon) = \sigma \sqrt{\frac{1 - e^{-2 \lambda \delta}}{2 \lambda}} 
\end{equation}

\noindent Solving for the parameters of the Ornstein-Uhlenbeck model yields

\begin{equation}
    \lambda = -\frac{ln(a)}{\delta}; \quad
    \mu = \frac{b}{1 - a}; \quad
    \sigma = sd(\epsilon) \sqrt{\frac{-2 ln(a)}{\delta (1 - a^2)}}
    \label{eq:OU_est}
\end{equation}

\noindent We therefore run the autoregressive model in equation (\ref{eq:linreg}) to obtain the $a$ and $b$ for each (i,j) pair, $a_{ij}$ and $b_{ij}$. Using equation (\ref{eq:OU_est}), the parameters $\mu_{ij}$, $\lambda_{ij}$, $\sigma_{ij}$ were estimated. These parameters allow the calculation of \textbf{Z-Scores}, which standardize deviations of the spread from its mean. The Z score of the spread can be computed using:

\begin{equation}
    Z = \frac{S_{t} - \mu}{\sigma}
    \label{eq:z_eqn}
\end{equation}

\subsection*{Signal Generation}
The Z-Score of the price spread determines trading signals:
\begin{itemize}
    \item \textbf{Enter Short Position:} If $Z > \text{Upper Threshold}$, short the spread (sell outperforming asset, buy underperforming asset).
    \item \textbf{Enter Long Position:} If $Z < \text{Lower Threshold}$, long the spread (buy underperforming asset, sell outperforming asset).
    \item \textbf{Exit Position:} If $Z$ reverts to within the exit thresholds or approaches zero, close the position.
\end{itemize}

\noindent In both strategies, once a z-score was computed, it was mapped to a percentile. The percentile is computed on a rolling basis, with a 90 day window as the lookback period to ensure that percentile values are compared with future values. This percentile was calculated as Point in time (PIT) based on the rolling window set in the function which was 90 days. In the case that the z-score fell above the 75th percentile, a short position on the spread would be taken the following day. If the z-score fell below the 25th percentile, a long position on the spread would be taken the following day. If the z-score crossed the 50th percentile, the position would be cleared. The performance of the pairs trading strategy was measured on the test data. For the strategy with the OU model, the autoregressive model was run on training data, and the performance was measured on the test data.

\section*{Results}

\begin{figure}[]
\centering
\includegraphics[width=0.8\textwidth]{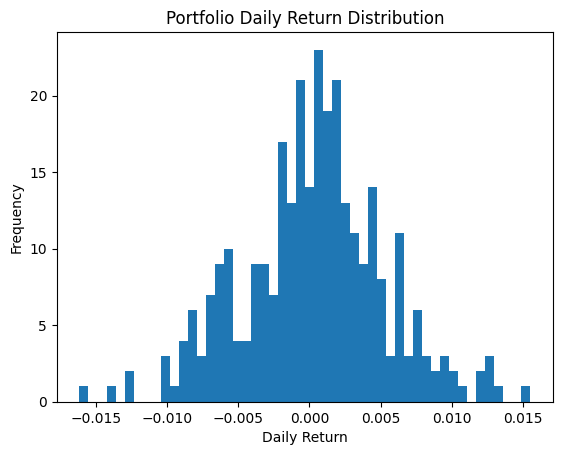}
\caption{Histogram Standard Model}
\label{fig:hist_standard}
\end{figure}

\begin{figure}[]
\centering
\includegraphics[width=0.8\textwidth]{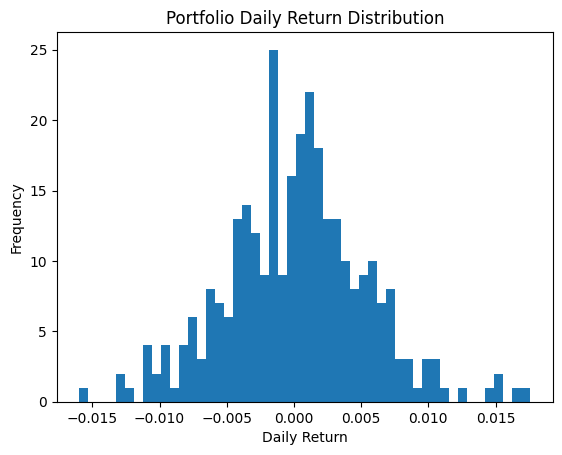}
\caption{Histogram OU Model}
\label{fig:ou_hist}
\end{figure}

\noindent Our preliminary results are shown in Table~\ref{tab:performance_metrics_standard} and Table~\ref{tab:performance_metrics_ou}. We find that the standard portfolio has an expected return of 7.21\% with a Sharpe ratio of 0.778. The frequency of the daily returns were also plotted and the results were shown in (\ref{fig:hist_standard}) and (\ref{fig:ou_hist}). \newline

\begin{figure}[]
\centering
\includegraphics[width=0.8\textwidth]{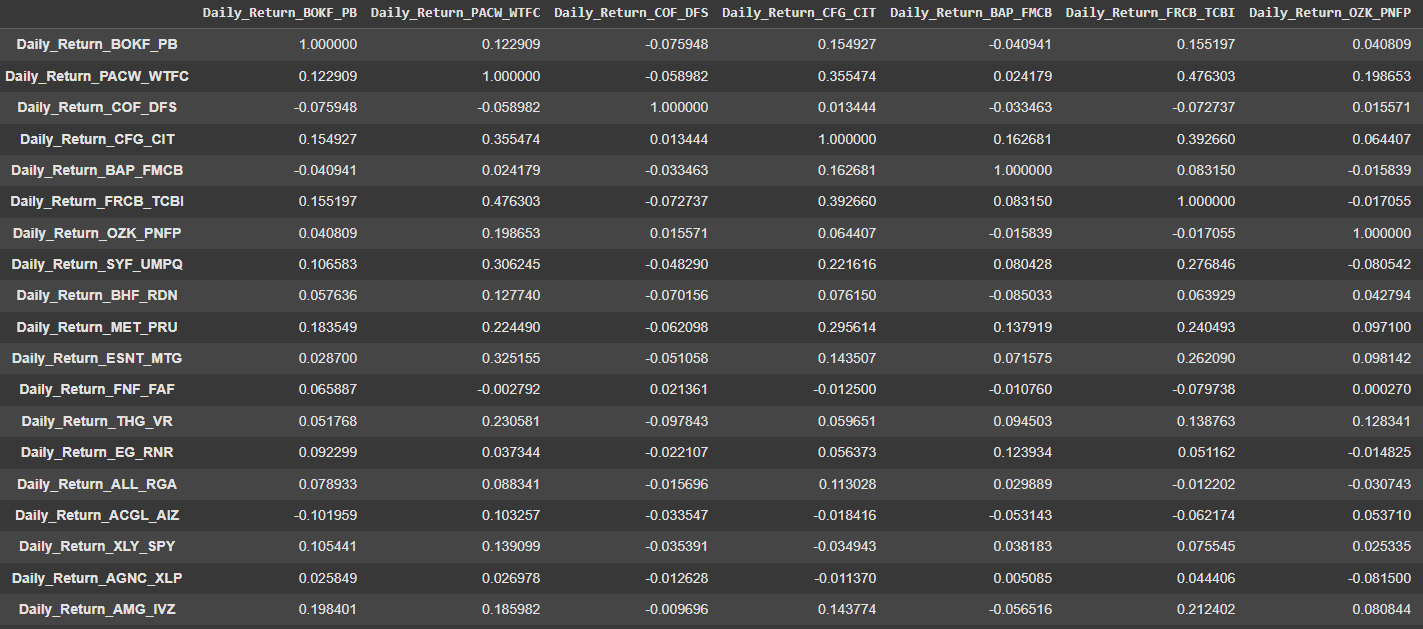}
\caption{Correlation Matrix Standard Model}
\label{fig:standard_correlation}
\end{figure}

\begin{figure}[]
\centering
\includegraphics[width=0.8\textwidth]{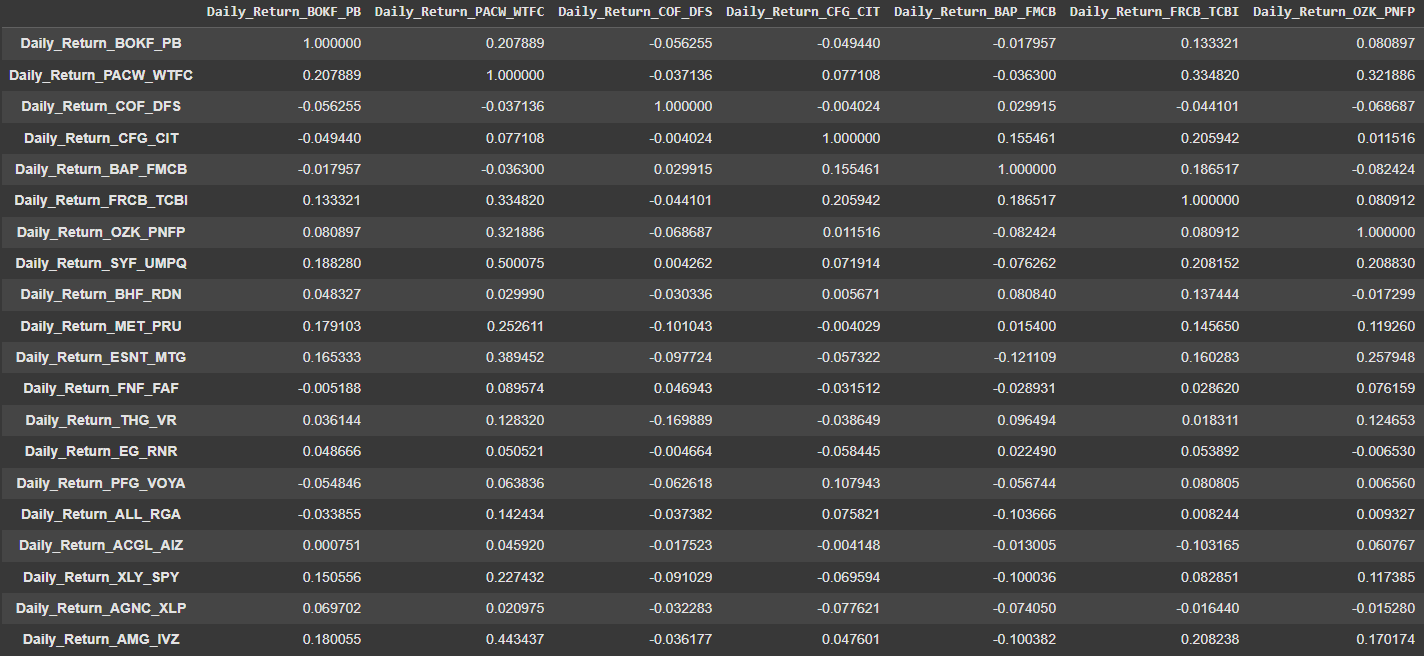}
\caption{Correlation Matrix OU Model}
\label{fig:ou_correlation}
\end{figure}

\noindent We find that the portfolio using the Ornstein-Uhlenbeck process has an expected return of 4.92\% with a Sharpe ratio of 0.468. Both strategies displayed a positive return. Examining the correlation matrix for both the standard and OU pairs trading strategy in Figures (\ref{fig:standard_correlation}) and \ref{fig:ou_correlation} shows that there is very little inter-pair correlation, which suggests that the returns are reasonably diversified and orthogonal to each other. \newline

\noindent Our findings show that the standard pairs trading strategy delivered a positive daily return 54.63 \% of the time and the OU pairs trading strategy deliver a positive daily return 53.3 \% of the time. While this is a strong win-loss ratio, further work must be done to test the robustness of the strategies on out of sample data to verify the robustness of the win-loss results.\newline

\begin{figure}[]
\centering
\includegraphics[width=0.8\textwidth]{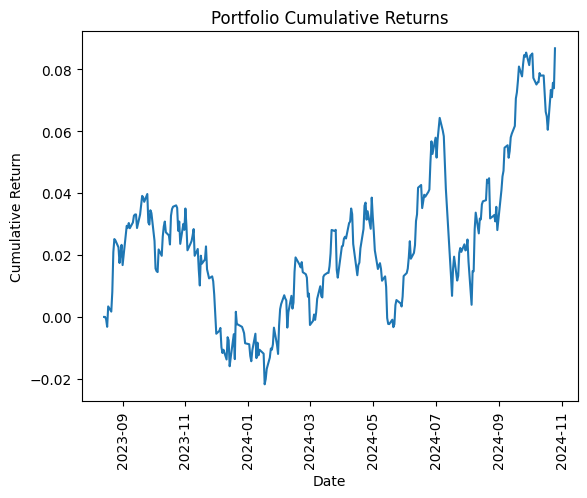}
\caption{Cumulative Returns Standard Model}
\label{fig:cumulative_standard}
\end{figure}

\begin{figure}[]
\centering
\includegraphics[width=0.8\textwidth]{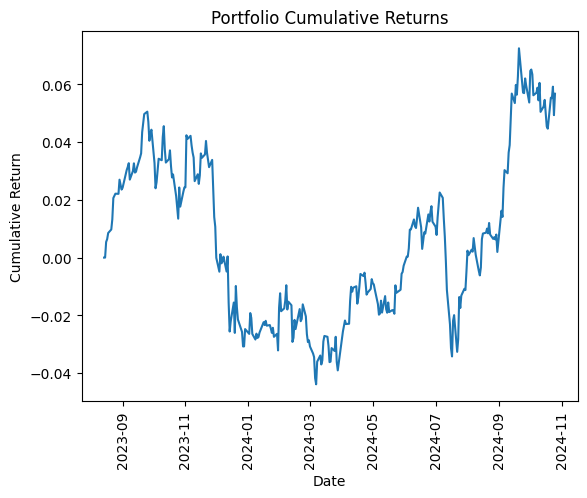}
\caption{Cumulative Returns OU Model}
\label{fig:cumulative_ou}
\end{figure}

\noindent Although the OU Model performed worse than the standard model, the cumulative returns plot indicates that the model captures the correct signals and overall trend, as shown in (\ref{fig:cumulative_standard}) and (\ref{fig:cumulative_ou}). This suggests that the model has strong predictive power but may have underperformed due to factors such as suboptimal parameter tuning or an inappropriate Z-score threshold. We suspect that this is likely due to potential non-stationarity of some pairs which led to poor mean reversion. To mitigate this, the Augmented Dickey-Fuller Test could have been used to test for stationarity of the pairs, which gives more reliability for the z-score calculation because the mean and standard deviation are constant. Nevertheless, further testing is required to ensure the robustness and validity of our results. Additional out of sample data and tests will need to be performed. Additionally, the cointegration test was performed in the period from 2018-2020 which was prior to the COVID-19 crisis. A change in regime may have led to a change in the behavior of the spread post-pandemic as well, and further work should be done to explore whether the price correlations were largely affected by major market regimes and events. 

\section*{Conclusion}

\noindent Our findings suggest that the pairs trading with Ornstein-Uhlenbeck process underperformed the naive pairs trading algorithm. The underperformance is likely due to the limited predictive power of the model and the fact that we are generalizing all the results to all companies. Future work can consider conditioning the pairs trading strategy on different industries and fine tuning parameters in the OU parameter estimation.

\section*{Discussion}

\noindent Future work should also explore different statistical methodologies to select pairs. Our pairs were chosen by choosing pairs that have the lowest mean squared distance between percentage price changes. Future work should explore different metrics or a combination such as Percentage Absolute difference, Correlation based distance, and Cosine Similarity to choose pairs. Testing the same pairs trading algorithm with fixed parameters while varying pair selection metrics can help evaluate the effectiveness of each metric. Incorporating additional autocorrelation tests may further refine pair selection.\newline

\noindent Additional work on portfolio optimization to maintain factor and/or sector neutrality could also be helpful. This could be done by finding additional time series of the returns of various factors or sectors, running a multivariate regression or principal component analysis (to further eliminate multicollinearity between factors) to solve for betas, and a quadratic program can be solved to optimize the weight of each spread in the portfolio to ensure that the portfolio respects factor and sector neutral constraints.\newline

\noindent Finally, hyperparameter tuning should also be explored to enhance the model's performance. Specifically, the Z-score thresholds, that is currently set arbitrarily at the 25th and 75th percentiles should be fine-tuned to account for variations in performances across different industries. 

\section*{Limitations}

\noindent The simulation did not incorporate transaction costs and if this strategy were to be implemented in real life the returns could be even lower. Slippage was also not accounted for. We assume that we transact on the next day and that the size of trades is negligible and does not move the market. The strategy also relies on the assumption that all the selected pairs exhibits mean-reverting behavior when in reality this assumption may not hold. Non-Stationarity pairs could lead to poor performance if the pairs fail to revert. Finally, the strategy was tested on a limited one-year time frame, which may not fully capture the effects of varying economic conditions such as booms, recessions, leadership changes, or global events like pandemics.

\newpage
\section*{Appendix}

\begin{table}[h!]
\centering
\begin{tabular}{|l|l|l|l|l|l|l|}
\hline
infocode & dscode  & isin         & ticker & dssecname     & region & marketdate \\ \hline
6347.0   & 151928  & US84265V1052 & SCCO   & SOUTHERN COPPER & US    & 2018-01-02 \\ \hline
6347.0   & 151928  & US84265V1052 & SCCO   & SOUTHERN COPPER & US    & 2018-01-03 \\ \hline
6347.0   & 151928  & US84265V1052 & SCCO   & SOUTHERN COPPER & US    & 2018-01-04 \\ \hline
6347.0   & 151928  & US84265V1052 & SCCO   & SOUTHERN COPPER & US    & 2018-01-05 \\ \hline
6347.0   & 151928  & US84265V1052 & SCCO   & SOUTHERN COPPER & US    & 2018-01-08 \\ \hline
\end{tabular}
\caption{Subsample of the dataset showing basic identifiers: infocode, dscode, isin, ticker, dssecname, region, and marketdate}
\label{tab:table1}
\end{table}

\begin{table}[h!]
\centering
\begin{tabular}{|l|l|l|l|}
\hline
marketdate & adjclose & general\_industry\_desc & industry\_group\_desc \\ \hline
2018-01-02 & 48.127152 & INDUSTRIAL & COPPER PRODUCERS \\ \hline
2018-01-03 & 48.215742 & INDUSTRIAL & COPPER PRODUCERS \\ \hline
2018-01-04 & 47.969693 & INDUSTRIAL & COPPER PRODUCERS \\ \hline
2018-01-05 & 48.343677 & INDUSTRIAL & COPPER PRODUCERS \\ \hline
2018-01-08 & 48.570053 & INDUSTRIAL & COPPER PRODUCERS \\ \hline
\end{tabular}
\caption{Subsample of the dataset showing financial and industry data: marketdate, adjclose, general\_industry\_desc, and industry\_group\_desc}
\label{tab:table2}
\end{table}

\begin{table}[h!]
\centering
\begin{tabular}{|l|c|}
\hline
\textbf{Metric}                     & \textbf{Value}      \\ \hline
Expected Return (Annualized)        & 0.072177            \\ \hline
Volatility (Annualized)             & 0.079848            \\ \hline
Sharpe Ratio                        & 0.778685            \\ \hline
Sortino Ratio                       & 1.202846            \\ \hline
Maximum Drawdown                    & -0.059133           \\ \hline
VaR (95\%)                          & -0.008078           \\ \hline
Win-Ratio                           & 0.546358            \\ \hline
\end{tabular}
\caption{Performance Metrics Summary for standard model}
\label{tab:performance_metrics_standard}
\end{table}

\begin{table}[]
\centering
\begin{tabular}{|l|c|}
\hline
\textbf{Metric}                     & \textbf{Value}      \\ \hline
Expected Return (Annualized)        & 0.049272            \\ \hline
Volatility (Annualized)             & 0.083920            \\ \hline
Sharpe Ratio                        & 0.467963            \\ \hline
Sortino Ratio                       & 0.773067            \\ \hline
Maximum Drawdown                    & -0.089884           \\ \hline
VaR (95\%)                          & -0.008473           \\ \hline
Win-Ratio                           & 0.533113            \\ \hline
\end{tabular}
\caption{Performance Metrics Summary OU Model}
\label{tab:performance_metrics_ou}
\end{table}

\begin{figure}[]
\centering
\includegraphics[width=0.8\textwidth]{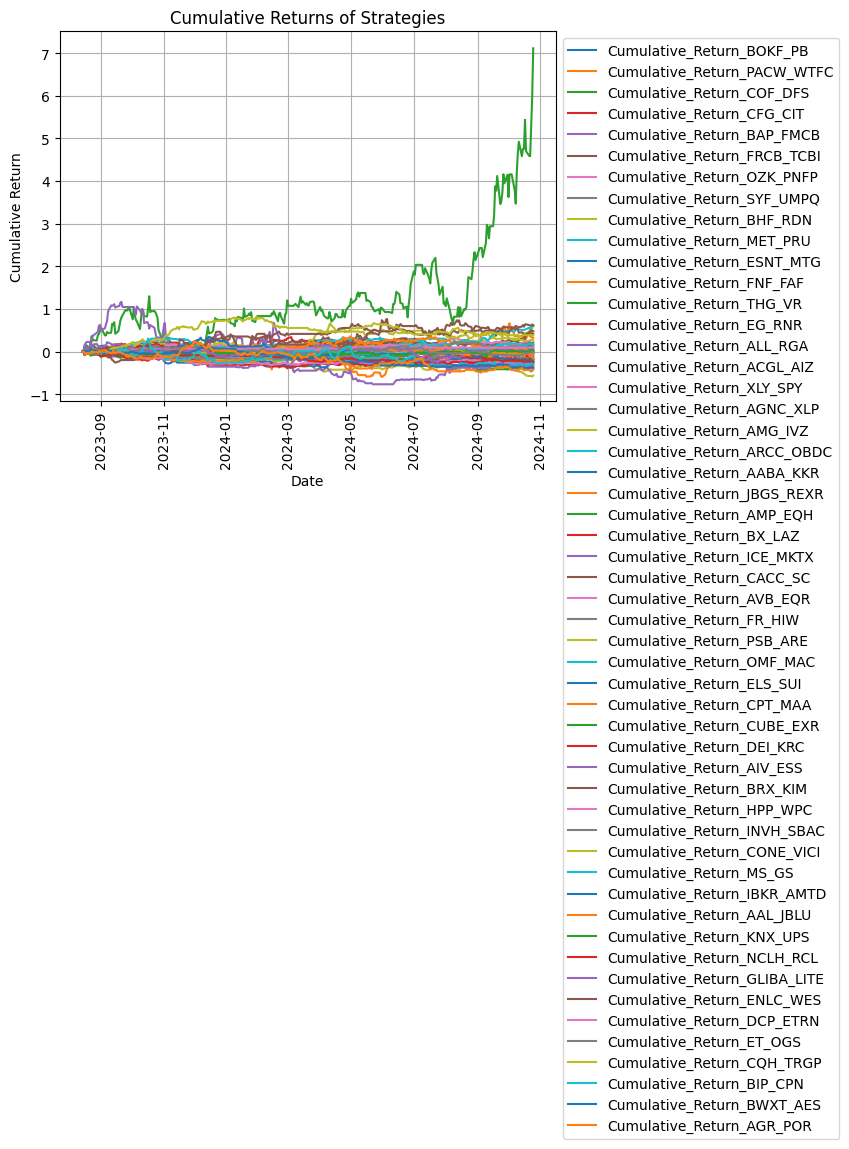}
\caption{Returns Standard Model}
\label{fig:returns_standard}
\end{figure}

\begin{figure}[]
\centering
\includegraphics[width=0.8\textwidth]{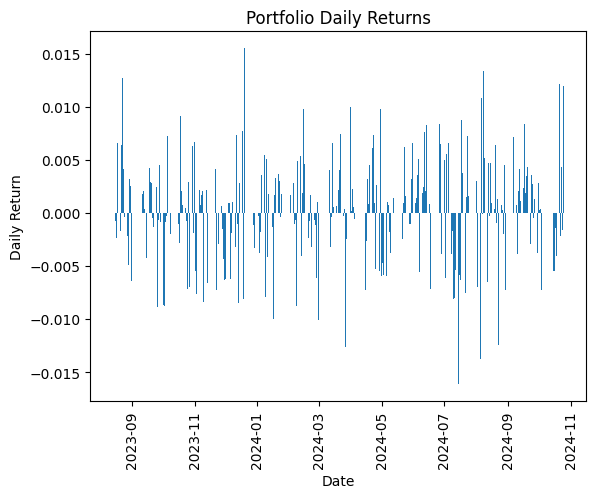}
\caption{Daily Returns Standard Model}
\label{fig:daily_standard}
\end{figure}

\begin{figure}[]
\centering
\includegraphics[width=0.8\textwidth]{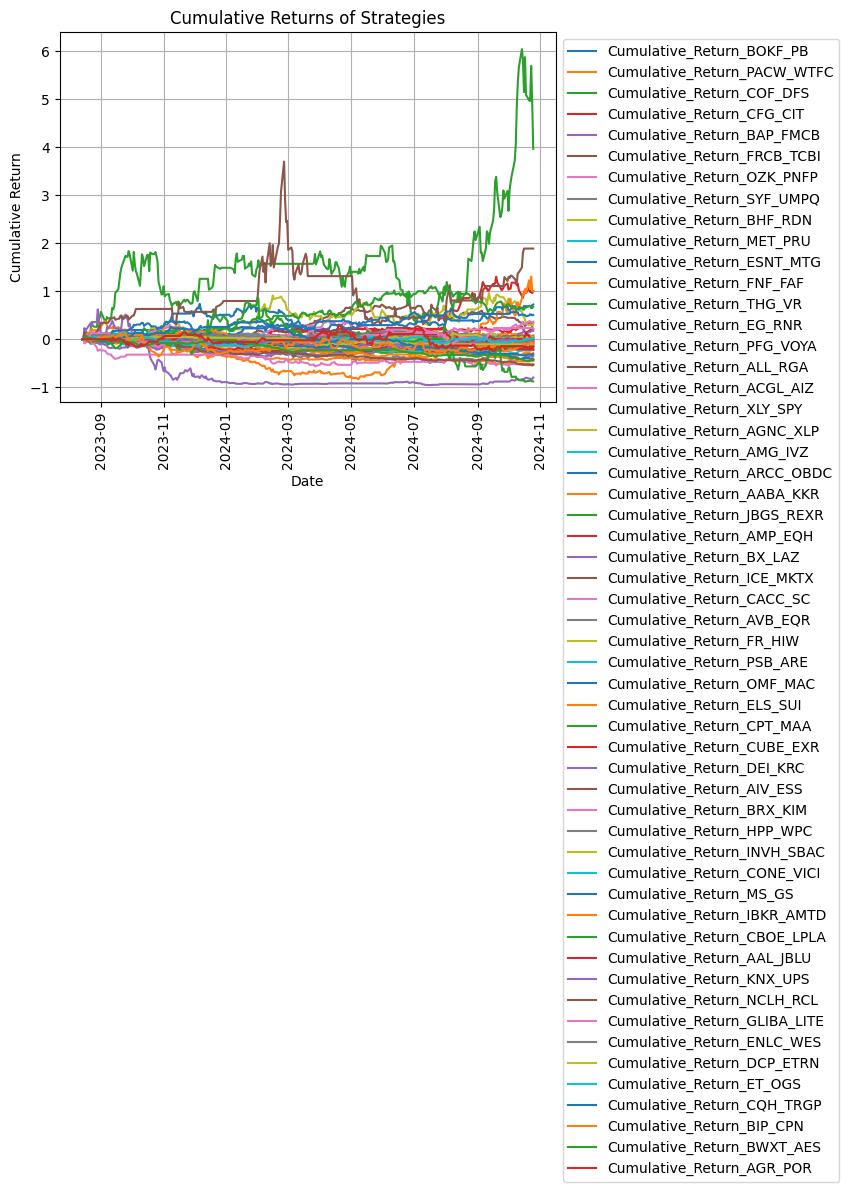}
\caption{Returns OU Model}
\label{fig:returns_ou}
\end{figure}

\begin{figure}[]
\centering
\includegraphics[width=0.8\textwidth]{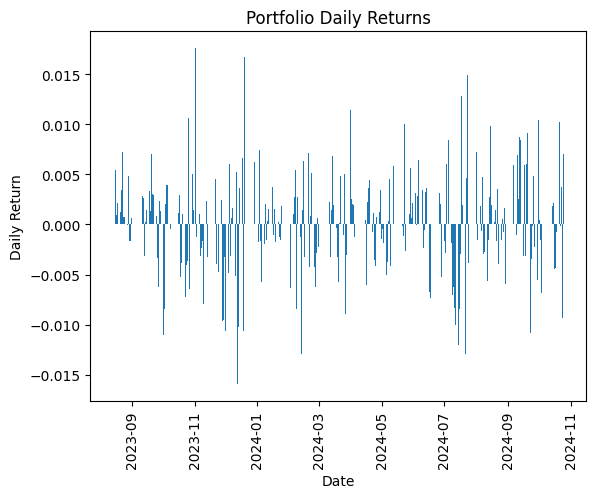}
\caption{Daily Returns OU Model}
\label{fig:daily_ou}
\end{figure}

\end{document}